\documentclass[aps,pra,twocolumn,superscriptaddress,nofootinbib]{revtex4-2}
\usepackage[utf8]{inputenc}
\usepackage{hyperref}
\usepackage{amsmath}
\usepackage{amssymb}
\usepackage{braket}
\usepackage{tikz, circuitikz, tikzit}
\usepackage{amsthm}

\usepackage{float}
\usepackage{comment}
\tikzstyle{env}=[copoint,regular polygon rotate=0,minimum width=0.2cm, fill=black]

\tikzstyle{probs}=[shape=semicircle,fill=white,draw=black,shape border rotate=180,minimum width=1.2cm]

\tikzstyle{nudge}=[yshift=0.6mm]

\tikzstyle{every picture}=[baseline=-0.25em,scale=0.5]
\tikzstyle{dotpic}=[] 
\tikzstyle{diredges}=[every to/.style={diredge}]
\tikzstyle{math matrix}=[matrix of math nodes,left delimiter=(,right delimiter=),inner sep=2pt,column sep=1em,row sep=0.5em,nodes={inner sep=0pt},text height=1.5ex, text depth=0.25ex]

\tikzstyle{inline text}=[text height=1.5ex, text depth=0.25ex,yshift=0.5mm]
\tikzstyle{label}=[font=\footnotesize,text height=1.5ex, text depth=0.25ex]
\tikzstyle{left label}=[label,anchor=east,xshift=2mm]
\tikzstyle{right label}=[label,anchor=west,xshift=-2mm]

\tikzstyle{braceedge}=[decorate,decoration={brace,amplitude=2mm,raise=-1mm}]
\tikzstyle{small braceedge}=[decorate,decoration={brace,amplitude=1mm,raise=-1mm}]

\tikzstyle{doubled}=[line width=1.6pt] 
\tikzstyle{boldedge}=[doubled,shorten <=-0.17mm,shorten >=-0.17mm]
\tikzstyle{boldedgegray}=[doubled,gray,shorten <=-0.17mm,shorten >=-0.17mm]
\tikzstyle{singleedgegray}=[gray]

\tikzstyle{semidoubled}=[line width=1.4pt] 
\tikzstyle{semiboldedgegray}=[semidoubled,gray,shorten <=-0.17mm,shorten >=-0.17mm]

\tikzstyle{boxedge}=[semiboldedgegray]

\tikzstyle{boldedgedashed}=[very thick,dashed,shorten <=-0.17mm,shorten >=-0.17mm]
\tikzstyle{vboldedgedashed}=[doubled,dashed,shorten <=-0.17mm,shorten >=-0.17mm]
\tikzstyle{left hook arrow}=[left hook-latex]
\tikzstyle{right hook arrow}=[right hook-latex]
\tikzstyle{sembracket}=[line width=0.5pt,shorten <=-0.07mm,shorten >=-0.07mm]

\tikzstyle{causal edge}=[->,thick,gray]
\tikzstyle{causal nondir}=[thick,gray]
\tikzstyle{timeline}=[thick,gray, dashed]

\tikzstyle{cedge}=[<->,thick,gray!70!white]

\tikzstyle{empty diagram}=[draw=gray!40!white,dashed,shape=rectangle,minimum width=1cm,minimum height=1cm]
\tikzstyle{empty diagram small}=[draw=gray!50!white,dashed,shape=rectangle,minimum width=0.6cm,minimum height=0.5cm]

\tikzstyle{dot}=[inner sep=0mm,minimum width=2mm,minimum height=2mm,draw,shape=circle]  
\tikzstyle{Wsquare}=[white dot, shape=regular polygon, rounded corners=0.8 mm, minimum size=3.3 mm, regular polygon sides=3, outer sep=-0.2mm]
\tikzstyle{Wsquareadj}=[white dot, shape=regular polygon, rounded corners=0.8 mm, minimum size=3.3 mm, regular polygon sides=3, outer sep=-0.2mm, regular polygon rotate=180]
\tikzstyle{ddot}=[inner sep=0mm, doubled, minimum width=2.5mm,minimum height=2.5mm,draw,shape=circle]

\tikzstyle{black dot}=[dot,fill=black]
\tikzstyle{white dot}=[dot,fill=white,,text depth=-0.2mm]
\tikzstyle{white Wsquare}=[Wsquare,fill=white,,text depth=-0.2mm]
\tikzstyle{white Wsquareadj}=[Wsquareadj,fill=white,,text depth=-0.2mm]
\tikzstyle{green dot}=[white dot] 
\tikzstyle{gray dot}=[dot,fill=gray!40!white,,text depth=-0.2mm]
\tikzstyle{red dot}=[gray dot] 

\tikzstyle{black ddot}=[ddot,fill=black]
\tikzstyle{white ddot}=[ddot,fill=white]
\tikzstyle{gray ddot}=[ddot,fill=gray!40!white]

\tikzstyle{gray edge}=[gray!60!white]

\tikzstyle{small dot}=[inner sep=0.5mm,minimum width=0pt,minimum height=0pt,draw,shape=circle]

\tikzstyle{small black dot}=[small dot,fill=black]
\tikzstyle{small white dot}=[small dot,fill=white]
\tikzstyle{small gray dot}=[small dot,fill=gray!40!white]

\tikzstyle{very small dot}=[inner sep=0.3mm,minimum width=0pt,minimum height=0pt,draw,shape=circle]

\tikzstyle{very small black dot}=[very small dot,fill=black]
\tikzstyle{very small white dot}=[small dot,fill=white]
\tikzstyle{very small gray dot}=[small dot,fill=gray!40!white]

\tikzstyle{causal dot}=[inner sep=0.4mm,minimum width=0pt,minimum height=0pt,draw=white,shape=circle,fill=gray!40!white]

\tikzstyle{phase dimensions}=[minimum size=5mm,font=\footnotesize,rectangle,rounded corners=2.5mm,inner sep=0.2mm,outer sep=-2mm]
\tikzstyle{dphase dimensions}=[minimum size=5mm,font=\footnotesize,rectangle,rounded corners=2.5mm,inner sep=0.2mm,outer sep=-2mm]

\tikzstyle{white phase dot}=[dot,fill=white,phase dimensions]
\tikzstyle{white phase ddot}=[ddot,fill=white,dphase dimensions]

\tikzstyle{white rect ddot}=[draw=black,fill=white,doubled,minimum size=5mm,font=\footnotesize,rectangle,rounded corners=2.5mm,inner sep=0.2mm]
\tikzstyle{gray rect ddot}=[draw=black,fill=gray!40!white,doubled,minimum size=6mm,font=\footnotesize,rectangle,rounded corners=3mm]

\tikzstyle{gray phase dot}=[dot,fill=gray!40!white,phase dimensions]
\tikzstyle{gray phase ddot}=[ddot,fill=gray!40!white,dphase dimensions]
\tikzstyle{grey phase dot}=[gray phase dot]
\tikzstyle{grey phase ddot}=[gray phase ddot]

\tikzstyle{small phase dimensions}=[minimum size=4mm,font=\tiny,rectangle,rounded corners=2mm,inner sep=0.2mm,outer sep=-2mm]
\tikzstyle{small dphase dimensions}=[minimum size=4mm,font=\tiny,rectangle,rounded corners=2mm,inner sep=0.2mm,outer sep=-2mm]

\tikzstyle{small gray phase dot}=[dot,fill=gray!40!white,small phase dimensions]
\tikzstyle{small gray phase ddot}=[ddot,fill=gray!40!white,small dphase dimensions]

\tikzstyle{small map}=[draw,shape=rectangle,minimum height=4mm,minimum width=4mm,fill=white]

\tikzstyle{cnot}=[fill=white,shape=circle,inner sep=-1.4pt]

\tikzstyle{asym hadamard}=[fill=white,draw,shape=NEbox,inner sep=0.6mm,font=\footnotesize,minimum height=4mm]
\tikzstyle{asym hadamard conj}=[fill=white,draw,shape=NWbox,inner sep=0.6mm,font=\footnotesize,minimum height=4mm]
\tikzstyle{asym hadamard dag}=[fill=white,draw,shape=SEbox,inner sep=0.6mm,font=\footnotesize,minimum height=4mm]

\tikzstyle{hadamard}=[fill=white,draw,inner sep=0.6mm,font=\footnotesize,minimum height=4mm,minimum width=4mm]
\tikzstyle{small hadamard}=[fill=white,draw,inner sep=0.6mm,minimum height=1.5mm,minimum width=1.5mm]
\tikzstyle{small hadamard rotate}=[small hadamard,rotate=45]
\tikzstyle{dhadamard}=[hadamard,doubled]
\tikzstyle{small dhadamard}=[small hadamard,doubled]
\tikzstyle{small dhadamard rotate}=[small hadamard rotate,doubled]
\tikzstyle{antipode}=[white dot,inner sep=0.3mm,font=\footnotesize]

\tikzstyle{scalar}=[diamond,draw,inner sep=0.5pt,font=\small]
\tikzstyle{dscalar}=[diamond,doubled, draw,inner sep=0.5pt,font=\small]

\tikzstyle{small box}=[rectangle,inline text,fill=white,draw,minimum height=5mm,yshift=-0.5mm,minimum width=5mm,font=\small]
\tikzstyle{small gray box}=[small box,fill=gray!30]
\tikzstyle{medium box}=[rectangle,inline text,fill=white,draw,minimum height=5mm,yshift=-0.5mm,minimum width=8mm,font=\small]
\tikzstyle{square box}=[small box] 
\tikzstyle{medium gray box}=[small box,fill=gray!30]
\tikzstyle{semilarge box}=[rectangle,inline text,fill=white,draw,minimum height=5mm,yshift=-0.5mm,minimum width=12.5mm,font=\small]
\tikzstyle{large box}=[rectangle,inline text,fill=white,draw,minimum height=5mm,yshift=-0.5mm,minimum width=15mm,font=\small]
\tikzstyle{large gray box}=[small box,fill=gray!30]

\tikzstyle{Bayes box}=[rectangle,fill=black,draw, minimum height=3mm, minimum width=3mm]

\tikzstyle{gray square point}=[small box,fill=gray!50]

\tikzstyle{dphase box white}=[dhadamard]
\tikzstyle{dphase box gray}=[dhadamard,fill=gray!50!white]
\tikzstyle{phase box white}=[hadamard]
\tikzstyle{phase box gray}=[hadamard,fill=gray!50!white]

\tikzstyle{point}=[regular polygon,regular polygon sides=3,draw,scale=0.75,inner sep=-0.5pt,minimum width=9mm,fill=white,regular polygon rotate=180]
\tikzstyle{copoint}=[regular polygon,regular polygon sides=3,draw,scale=0.75,inner sep=-0.5pt,minimum width=9mm,fill=white]
\tikzstyle{dpoint}=[point,doubled]
\tikzstyle{dcopoint}=[copoint,doubled]

\tikzstyle{wide copoint}=[fill=white,draw,shape=isosceles triangle,shape border rotate=90,isosceles triangle stretches=true,inner sep=0pt,minimum width=1.5cm,minimum height=6.12mm]
\tikzstyle{wide point}=[fill=white,draw,shape=isosceles triangle,shape border rotate=-90,isosceles triangle stretches=true,inner sep=0pt,minimum width=1.5cm,minimum height=6.12mm,yshift=-0.0mm]
\tikzstyle{wide point plus}=[fill=white,draw,shape=isosceles triangle,shape border rotate=-90,isosceles triangle stretches=true,inner sep=0pt,minimum width=1.74cm,minimum height=7mm,yshift=-0.0mm]

\tikzstyle{wide dpoint}=[fill=white,doubled,draw,shape=isosceles triangle,shape border rotate=-90,isosceles triangle stretches=true,inner sep=0pt,minimum width=1.5cm,minimum height=6.12mm,yshift=-0.0mm]

\tikzstyle{tinypoint}=[regular polygon,regular polygon sides=3,draw,scale=0.55,inner sep=-0.15pt,minimum width=6mm,fill=white,regular polygon rotate=180] 

\tikzstyle{white point}=[point]
\tikzstyle{white dpoint}=[dpoint]
\tikzstyle{green point}=[white point] 
\tikzstyle{white copoint}=[copoint]
\tikzstyle{gray point}=[point,fill=gray!40!white]
\tikzstyle{gray dpoint}=[gray point,doubled]
\tikzstyle{red point}=[gray point] 
\tikzstyle{gray copoint}=[copoint,fill=gray!40!white]
\tikzstyle{gray dcopoint}=[gray copoint,doubled]

\tikzstyle{white point guide}=[regular polygon,regular polygon sides=3,font=\scriptsize,draw,scale=0.65,inner sep=-0.5pt,minimum width=9mm,fill=white,regular polygon rotate=180]

\tikzstyle{black point}=[point,fill=black,font=\color{white}]
\tikzstyle{black copoint}=[copoint,fill=black,font=\color{white}]

\tikzstyle{tiny gray point}=[tinypoint,fill=gray!40!white]

\tikzstyle{diredge}=[->]
\tikzstyle{ddiredge}=[<->]
\tikzstyle{rdiredge}=[<-]
\tikzstyle{thickdiredge}=[->, very thick]
\tikzstyle{pointer edge}=[->,very thick,gray]
\tikzstyle{pointer edge part}=[very thick,gray]
\tikzstyle{dashed edge}=[dashed]
\tikzstyle{thick dashed edge}=[very thick,dashed]
\tikzstyle{thick gray dashed edge}=[thick dashed edge,gray!40]
\tikzstyle{thick map edge}=[very thick,|->]

\makeatletter
\newcommand{\boxshape}[3]{%
\pgfdeclareshape{#1}{
\inheritsavedanchors[from=rectangle] 
\inheritanchorborder[from=rectangle]
\inheritanchor[from=rectangle]{center}
\inheritanchor[from=rectangle]{north}
\inheritanchor[from=rectangle]{south}
\inheritanchor[from=rectangle]{west}
\inheritanchor[from=rectangle]{east}
\backgroundpath{
\southwest \pgf@xa=\pgf@x \pgf@ya=\pgf@y
\northeast \pgf@xb=\pgf@x \pgf@yb=\pgf@y

\@tempdima=#2
\@tempdimb=#3

\pgfpathmoveto{\pgfpoint{\pgf@xa - 5pt + \@tempdima}{\pgf@ya}}
\pgfpathlineto{\pgfpoint{\pgf@xa - 5pt - \@tempdima}{\pgf@yb}}
\pgfpathlineto{\pgfpoint{\pgf@xb + 5pt + \@tempdimb}{\pgf@yb}}
\pgfpathlineto{\pgfpoint{\pgf@xb + 5pt - \@tempdimb}{\pgf@ya}}
\pgfpathlineto{\pgfpoint{\pgf@xa - 5pt + \@tempdima}{\pgf@ya}}
\pgfpathclose
}
}}

\boxshape{NEbox}{0pt}{0pt}
\boxshape{SEbox}{0pt}{-5pt}
\boxshape{NWbox}{5pt}{0pt}
\boxshape{SWbox}{-5pt}{0pt}
\boxshape{EBox}{-3pt}{3pt}
\boxshape{WBox}{3pt}{-3pt}
\makeatother

\tikzstyle{cloud}=[shape=cloud,draw,minimum width=1.5cm,minimum height=1.5cm]

\tikzstyle{map}=[draw,shape=NEbox,inner sep=2pt,minimum height=6mm,fill=white]
\tikzstyle{dashedmap}=[draw,dashed,gray,shape=NEbox,inner sep=2pt,minimum height=6mm,fill=white]

\tikzstyle{dashed map}=[fill=white, draw=gray, shape=rectangle, style=map, dashed]

\tikzstyle{mapdag}=[draw,shape=SEbox,inner sep=2pt,minimum height=6mm,fill=white]
\tikzstyle{mapadj}=[draw,shape=SEbox,inner sep=2pt,minimum height=6mm,fill=white]
\tikzstyle{maptrans}=[draw,shape=SWbox,inner sep=2pt,minimum height=6mm,fill=white]
\tikzstyle{mapconj}=[draw,shape=NWbox,inner sep=2pt,minimum height=6mm,fill=white]

\tikzstyle{medium map}=[draw,shape=NEbox,inner sep=2pt,minimum height=6mm,fill=white,minimum width=7mm]
\tikzstyle{medium map dag}=[draw,shape=SEbox,inner sep=2pt,minimum height=6mm,fill=white,minimum width=7mm]
\tikzstyle{medium map adj}=[draw,shape=SEbox,inner sep=2pt,minimum height=6mm,fill=white,minimum width=7mm]
\tikzstyle{medium map trans}=[draw,shape=SWbox,inner sep=2pt,minimum height=6mm,fill=white,minimum width=7mm]
\tikzstyle{medium map conj}=[draw,shape=NWbox,inner sep=2pt,minimum height=6mm,fill=white,minimum width=7mm]
\tikzstyle{semilarge map}=[draw,shape=NEbox,inner sep=2pt,minimum height=6mm,fill=white,minimum width=9.5mm]
\tikzstyle{semilarge map trans}=[draw,shape=SWbox,inner sep=2pt,minimum height=6mm,fill=white,minimum width=9.5mm]
\tikzstyle{semilarge map adj}=[draw,shape=SEbox,inner sep=2pt,minimum height=6mm,fill=white,minimum width=9.5mm]
\tikzstyle{semilarge map dag}=[draw,shape=SEbox,inner sep=2pt,minimum height=6mm,fill=white,minimum width=9.5mm]
\tikzstyle{semilarge map conj}=[draw,shape=NWbox,inner sep=2pt,minimum height=6mm,fill=white,minimum width=9.5mm]
\tikzstyle{large map}=[draw,shape=NEbox,inner sep=2pt,minimum height=6mm,fill=white,minimum width=12mm]
\tikzstyle{large map conj}=[draw,shape=NWbox,inner sep=2pt,minimum height=6mm,fill=white,minimum width=12mm]
\tikzstyle{very large map}=[draw,shape=NEbox,inner sep=2pt,minimum height=6mm,fill=white,minimum width=17mm]
\tikzstyle{very very large map}=[draw,shape=NEbox,inner sep=2pt,minimum height=6mm,fill=white,minimum width=50mm]
\tikzstyle{large map dag}=[draw,shape=SEbox,inner sep=2pt,minimum height=6mm,fill=white,minimum width=12mm]

\tikzstyle{medium dmap}=[draw,doubled,shape=NEbox,inner sep=2pt,minimum height=6mm,fill=white,minimum width=7mm]
\tikzstyle{medium dmap dag}=[draw,doubled,shape=SEbox,inner sep=2pt,minimum height=6mm,fill=white,minimum width=7mm]
\tikzstyle{medium dmap adj}=[draw,doubled,shape=SEbox,inner sep=2pt,minimum height=6mm,fill=white,minimum width=7mm]
\tikzstyle{medium dmap trans}=[draw,doubled,shape=SWbox,inner sep=2pt,minimum height=6mm,fill=white,minimum width=7mm]
\tikzstyle{medium dmap conj}=[draw,doubled,shape=NWbox,inner sep=2pt,minimum height=6mm,fill=white,minimum width=7mm]
\tikzstyle{semilarge dmap}=[draw,doubled,shape=NEbox,inner sep=2pt,minimum height=6mm,fill=white,minimum width=9.5mm]
\tikzstyle{semilarge dmap trans}=[draw,doubled,shape=SWbox,inner sep=2pt,minimum height=6mm,fill=white,minimum width=9.5mm]
\tikzstyle{semilarge dmap adj}=[draw,doubled,shape=SEbox,inner sep=2pt,minimum height=6mm,fill=white,minimum width=9.5mm]
\tikzstyle{semilarge dmap dag}=[draw,doubled,shape=SEbox,inner sep=2pt,minimum height=6mm,fill=white,minimum width=9.5mm]
\tikzstyle{semilarge dmap conj}=[draw,doubled,shape=NWbox,inner sep=2pt,minimum height=6mm,fill=white,minimum width=9.5mm]
\tikzstyle{large dmap}=[draw,doubled,shape=NEbox,inner sep=2pt,minimum height=6mm,fill=white,minimum width=12mm]
\tikzstyle{large dmap conj}=[draw,doubled,shape=NWbox,inner sep=2pt,minimum height=6mm,fill=white,minimum width=12mm]
\tikzstyle{large dmap trans}=[draw,doubled,shape=SWbox,inner sep=2pt,minimum height=6mm,fill=white,minimum width=12mm]
\tikzstyle{large dmap adj}=[draw,doubled,shape=SEbox,inner sep=2pt,minimum height=6mm,fill=white,minimum width=12mm]
\tikzstyle{large dmap dag}=[draw,doubled,shape=SEbox,inner sep=2pt,minimum height=6mm,fill=white,minimum width=12mm]
\tikzstyle{very large dmap}=[draw,doubled,shape=NEbox,inner sep=2pt,minimum height=6mm,fill=white,minimum width=19.5mm]

\tikzstyle{muxbox}=[draw,shape=rectangle,minimum height=3mm,minimum width=3mm,fill=white]
\tikzstyle{dmuxbox}=[muxbox,doubled]

\tikzstyle{box}=[draw,shape=rectangle,inner sep=2pt,minimum height=6mm,minimum width=6mm,fill=white]
\tikzstyle{dbox}=[draw,doubled,shape=rectangle,inner sep=2pt,minimum height=6mm,minimum width=6mm,fill=white]
\tikzstyle{dmap}=[draw,doubled,shape=NEbox,inner sep=2pt,minimum height=6mm,fill=white]
\tikzstyle{dmapdag}=[draw,doubled,shape=SEbox,inner sep=2pt,minimum height=6mm,fill=white]
\tikzstyle{dmapadj}=[draw,doubled,shape=SEbox,inner sep=2pt,minimum height=6mm,fill=white]
\tikzstyle{dmaptrans}=[draw,doubled,shape=SWbox,inner sep=2pt,minimum height=6mm,fill=white]
\tikzstyle{dmapconj}=[draw,doubled,shape=NWbox,inner sep=2pt,minimum height=6mm,fill=white]

\tikzstyle{ddmap}=[draw,doubled,dashed,shape=NEbox,inner sep=2pt,minimum height=6mm,fill=white]
\tikzstyle{ddmapdag}=[draw,doubled,dashed,shape=SEbox,inner sep=2pt,minimum height=6mm,fill=white]
\tikzstyle{ddmapadj}=[draw,doubled,dashed,shape=SEbox,inner sep=2pt,minimum height=6mm,fill=white]
\tikzstyle{ddmaptrans}=[draw,doubled,dashed,shape=SWbox,inner sep=2pt,minimum height=6mm,fill=white]
\tikzstyle{ddmapconj}=[draw,doubled,dashed,shape=NWbox,inner sep=2pt,minimum height=6mm,fill=white]

\boxshape{sNEbox}{0pt}{3pt}
\boxshape{sSEbox}{0pt}{-3pt}
\boxshape{sNWbox}{3pt}{0pt}
\boxshape{sSWbox}{-3pt}{0pt}
\tikzstyle{smap}=[draw,shape=sNEbox,fill=white]
\tikzstyle{smapdag}=[draw,shape=sSEbox,fill=white]
\tikzstyle{smapadj}=[draw,shape=sSEbox,fill=white]
\tikzstyle{smaptrans}=[draw,shape=sSWbox,fill=white]
\tikzstyle{smapconj}=[draw,shape=sNWbox,fill=white]

\tikzstyle{dsmap}=[draw,dashed,shape=sNEbox,fill=white]
\tikzstyle{dsmapdag}=[draw,dashed,shape=sSEbox,fill=white]
\tikzstyle{dsmaptrans}=[draw,dashed,shape=sSWbox,fill=white]
\tikzstyle{dsmapconj}=[draw,dashed,shape=sNWbox,fill=white]

\boxshape{mNEbox}{0pt}{10pt}
\boxshape{mSEbox}{0pt}{-10pt}
\boxshape{mNWbox}{10pt}{0pt}
\boxshape{mSWbox}{-10pt}{0pt}
\tikzstyle{mmap}=[draw,shape=mNEbox]
\tikzstyle{mmapdag}=[draw,shape=mSEbox]
\tikzstyle{mmaptrans}=[draw,shape=mSWbox]
\tikzstyle{mmapconj}=[draw,shape=mNWbox]

\tikzstyle{mmapgray}=[draw,fill=gray!40!white,shape=mNEbox]
\tikzstyle{smapgray}=[draw,fill=gray!40!white,shape=sNEbox]

\makeatletter

\pgfdeclareshape{cornerpoint}{
\inheritsavedanchors[from=rectangle] 
\inheritanchorborder[from=rectangle]
\inheritanchor[from=rectangle]{center}
\inheritanchor[from=rectangle]{north}
\inheritanchor[from=rectangle]{south}
\inheritanchor[from=rectangle]{west}
\inheritanchor[from=rectangle]{east}
\backgroundpath{
\southwest \pgf@xa=\pgf@x \pgf@ya=\pgf@y
\northeast \pgf@xb=\pgf@x \pgf@yb=\pgf@y

\pgfmathsetmacro{\pgf@shorten@left}{\pgfkeysvalueof{/tikz/shorten left}}
\pgfmathsetmacro{\pgf@shorten@right}{\pgfkeysvalueof{/tikz/shorten right}}

\pgfpathmoveto{\pgfpoint{0.5 * (\pgf@xa + \pgf@xb)}{\pgf@ya - 5pt}}
\pgfpathlineto{\pgfpoint{\pgf@xa - 8pt + \pgf@shorten@left}{\pgf@yb - 1.5 * \pgf@shorten@left}}
\pgfpathlineto{\pgfpoint{\pgf@xa - 8pt + \pgf@shorten@left}{\pgf@yb}}
\pgfpathlineto{\pgfpoint{\pgf@xb + 8pt - \pgf@shorten@right}{\pgf@yb}}
\pgfpathlineto{\pgfpoint{\pgf@xb + 8pt - \pgf@shorten@right}{\pgf@yb - 1.5 * \pgf@shorten@right}}
\pgfpathclose
}
}

\pgfdeclareshape{cornercopoint}{
\inheritsavedanchors[from=rectangle] 
\inheritanchorborder[from=rectangle]
\inheritanchor[from=rectangle]{center}
\inheritanchor[from=rectangle]{north}
\inheritanchor[from=rectangle]{south}
\inheritanchor[from=rectangle]{west}
\inheritanchor[from=rectangle]{east}
\backgroundpath{
\southwest \pgf@xa=\pgf@x \pgf@ya=\pgf@y
\northeast \pgf@xb=\pgf@x \pgf@yb=\pgf@y

\pgfmathsetmacro{\pgf@shorten@left}{\pgfkeysvalueof{/tikz/shorten left}}
\pgfmathsetmacro{\pgf@shorten@right}{\pgfkeysvalueof{/tikz/shorten right}}

\pgfpathmoveto{\pgfpoint{0.5 * (\pgf@xa + \pgf@xb)}{\pgf@yb + 5pt}}
\pgfpathlineto{\pgfpoint{\pgf@xa - 8pt + \pgf@shorten@left}{\pgf@ya + 1.5 * \pgf@shorten@left}}
\pgfpathlineto{\pgfpoint{\pgf@xa - 8pt + \pgf@shorten@left}{\pgf@ya}}
\pgfpathlineto{\pgfpoint{\pgf@xb + 8pt - \pgf@shorten@right}{\pgf@ya}}
\pgfpathlineto{\pgfpoint{\pgf@xb + 8pt - \pgf@shorten@right}{\pgf@ya + 1.5 * \pgf@shorten@right}}
\pgfpathclose
}
}

\makeatother

\pgfkeyssetvalue{/tikz/shorten left}{0pt}
\pgfkeyssetvalue{/tikz/shorten right}{0pt}

\tikzstyle{kpoint common}=[draw,fill=white,inner sep=1pt,minimum height=4mm]
\tikzstyle{kpoint sc}=[shape=cornerpoint,kpoint common]
\tikzstyle{kpoint adjoint sc}=[shape=cornercopoint,kpoint common]
\tikzstyle{kpoint}=[shape=cornerpoint,shorten left=5pt,kpoint common]
\tikzstyle{kpoint adjoint}=[shape=cornercopoint,shorten left=5pt,kpoint common]
\tikzstyle{kpoint conjugate}=[shape=cornerpoint,shorten right=5pt,kpoint common]
\tikzstyle{kpoint transpose}=[shape=cornercopoint,shorten right=5pt,kpoint common]
\tikzstyle{kpoint symm}=[shape=cornerpoint,shorten left=5pt,shorten right=5pt,kpoint common]

\tikzstyle{black kpoint}=[shape=cornerpoint,shorten left=5pt,kpoint common,fill=black,font=\color{white}]
\tikzstyle{black kpoint adjoint}=[shape=cornercopoint,shorten left=5pt,kpoint common,fill=black,font=\color{white}]
\tikzstyle{black kpointadj}=[shape=cornercopoint,shorten left=5pt,kpoint common,fill=black,font=\color{white}]

\tikzstyle{black dkpoint}=[shape=cornerpoint,shorten left=5pt,kpoint common,fill=black, doubled,font=\color{white}]
\tikzstyle{black dkpoint adjoint}=[shape=cornercopoint,shorten left=5pt,kpoint common,fill=black, doubled,font=\color{white}]
\tikzstyle{black dkpointadj}=[shape=cornercopoint,shorten left=5pt,kpoint common,fill=black, doubled,font=\color{white}] 

\tikzstyle{kpointdag}=[kpoint adjoint]
\tikzstyle{kpointadj}=[kpoint adjoint]
\tikzstyle{kpointconj}=[kpoint conjugate]
\tikzstyle{kpointtrans}=[kpoint transpose]

\tikzstyle{big kpoint}=[kpoint, minimum width=1.2 cm, minimum height=8mm, inner sep=4pt, text depth=3mm]

\tikzstyle{wide kpoint}=[kpoint, minimum width=1 cm, inner sep=2pt]
\tikzstyle{wide kpointdag}=[kpointdag, minimum width=1 cm, inner sep=2pt]
\tikzstyle{wide kpointconj}=[kpointconj, minimum width=1 cm, inner sep=2pt]
\tikzstyle{wide kpointtrans}=[kpointtrans, minimum width=1 cm, inner sep=2pt]

\tikzstyle{gray kpoint}=[kpoint,fill=gray!50!white]
\tikzstyle{gray kpointdag}=[kpointdag,fill=gray!50!white]
\tikzstyle{gray kpointadj}=[kpointadj,fill=gray!50!white]
\tikzstyle{gray kpointconj}=[kpointconj,fill=gray!50!white]
\tikzstyle{gray kpointtrans}=[kpointtrans,fill=gray!50!white]

\tikzstyle{gray dkpoint}=[kpoint,fill=gray!50!white,doubled]
\tikzstyle{gray dkpointdag}=[kpointdag,fill=gray!50!white,doubled]
\tikzstyle{gray dkpointadj}=[kpointadj,fill=gray!50!white,doubled]
\tikzstyle{gray dkpointconj}=[kpointconj,fill=gray!50!white,doubled]
\tikzstyle{gray dkpointtrans}=[kpointtrans,fill=gray!50!white,doubled]

\tikzstyle{white label}=[draw,fill=white,rectangle,inner sep=0.7 mm]
\tikzstyle{gray label}=[draw,fill=gray!50!white,rectangle,inner sep=0.7 mm]
\tikzstyle{black label}=[draw,fill=black,rectangle,inner sep=0.7 mm]

\tikzstyle{dkpoint}=[kpoint,doubled]
\tikzstyle{wide dkpoint}=[wide kpoint,doubled]
\tikzstyle{dkpointdag}=[kpoint adjoint,doubled]
\tikzstyle{wide dkpointdag}=[wide kpointdag,doubled]
\tikzstyle{dkcopoint}=[kpoint adjoint,doubled]
\tikzstyle{dkpointadj}=[kpoint adjoint,doubled]
\tikzstyle{dkpointconj}=[kpoint conjugate,doubled]
\tikzstyle{dkpointtrans}=[kpoint transpose,doubled]

\tikzstyle{kscalar}=[kpoint common, shape=EBox, inner xsep=-1pt, inner ysep=3pt,font=\small]
\tikzstyle{kscalarconj}=[kpoint common, shape=WBox, inner xsep=-1pt, inner ysep=3pt,font=\small]

\tikzstyle{spekpoint}=[kpoint sc,minimum height=5mm,inner sep=3pt]
\tikzstyle{spekcopoint}=[kpoint adjoint sc,minimum height=5mm,inner sep=3pt]

\tikzstyle{dspekpoint}=[spekpoint,doubled]
\tikzstyle{dspekcopoint}=[spekcopoint,doubled]

 \tikzstyle{discard}=[circuit ee IEC, ground,rotate=180,scale=1.5,inner sep=-2mm]
 \tikzstyle{downground}=[circuit ee IEC,thick,ground,rotate=-90,scale=1.5,inner sep=-2mm]

\tikzstyle{maxmix}=[regular polygon,regular polygon sides=3,draw=black,xscale=0.4,yscale=0.3,inner sep=-0.5pt,minimum width=10mm,fill=gray,regular polygon rotate=180]

 \tikzstyle{bigground}=[regular polygon,regular polygon sides=3,draw=gray,scale=0.50,inner sep=-0.5pt,minimum width=10mm,fill=gray]

\tikzstyle{arrs}=[-latex,font=\small,auto]
\tikzstyle{arrow plain}=[arrs]
\tikzstyle{arrow dashed}=[dashed,arrs]
\tikzstyle{arrow bold}=[very thick,arrs]
\tikzstyle{arrow hide}=[draw=white!0,-]
\tikzstyle{arrow reverse}=[latex-]
\tikzstyle{cdnode}=[]

\tikzstyle{green dashed arrow}=[green, arrow dashed]
\tikzstyle{red dashed arrow}=[red, arrow dashed]

\usetikzlibrary{backgrounds}
\usetikzlibrary{arrows}
\usetikzlibrary{shapes,shapes.geometric,shapes.misc}
\usetikzlibrary{positioning,chains,fit,shapes,calc}
\usepackage[shortlabels]{enumitem}

\tikzstyle{tikzfig}=[baseline=-0.25em,scale=4]

\pgfkeys{/tikz/tikzit fill/.initial=0}
\pgfkeys{/tikz/tikzit draw/.initial=0}
\pgfkeys{/tikz/tikzit shape/.initial=0}
\pgfkeys{/tikz/tikzit category/.initial=0}

\pgfdeclarelayer{edgelayer}
\pgfdeclarelayer{nodelayer}
\pgfsetlayers{background,edgelayer,nodelayer,main}

\tikzstyle{none}=[inner sep=0mm]

\tikzstyle{every loop}=[]

\def\be{\begin{equation}}
\def\ee{\end{equation}}
\def\ba{\begin{align}}
\def\ea{\end{align}}

\newtheorem{definition}{Definition}

\newtheorem{theorem}{Theorem}

\newtheorem*{theorem*}{Theorem}

\begin{document}
\title{A no-go theorem for absolute observed events without inequalities or modal logic}

\author{Nick Ormrod}
\email{nicholas.ormrod@cs.ox.ac.uk}
\affiliation{Quantum Group, Department of Computer Science, University of Oxford}

\author{Jonathan Barrett}
\email{jonathan.barrett@cs.ox.ac.uk}
\affiliation{Quantum Group, Department of Computer Science, University of Oxford}

\begin{abstract}
    This paper builds on no-go theorems to the effect that quantum theory is inconsistent with observations being \textit{absolute}; that is, unique and non-relative. 
    Unlike the existing no-go results, the one introduced here is based on a theory-independent absoluteness assumption, \textit{and} there is no need to assume the validity of standard probability theory or of modal logic.
    The contradiction is derived by assuming that quantum theory applies in any inertial reference frame; accordingly, the result also illuminates a tension between \textit{special relativity} and absoluteness.
\end{abstract}

\maketitle

\section{Introduction}

It is standard practice in quantum theory to assume that the state of a closed system evolves unitarily right up until the moment of measurement. At that point, it is supposed to discontinuously `collapse' to a new state corresponding to one of the possible outcomes, with a probability given by the Born rule. 
This leads to an ambiguity \cite{Baumann_2018} when we consider a \textit{supermeasurement} \cite{wigner1995remarks}; that is, a measurement on a system that includes an observer who has just performed her own measurement.
When we calculate the probabilities for the supermeasurement, should we assume that all evolution prior to the supermeasurement was unitary, including the initial measurement? 
Or should we assume that the initial measurement collapsed the state?

The second approach involves admitting that `measurements' (however they are defined) taking place within closed systems must be treated differently to the processes in closed systems that have so far been successfully modelled with quantum theory. If, on the other hand, one wants to maintain that the laws governing small systems are the same as, and not mere approximations to, the laws governing systems in which `measurements' can take place, then one will have to model the initial measurement unitarily. 

But it is not clear that it even \textit{makes sense} to model measurements unitarily. For a unitary description of a measurement often leads to a state in which the agent's observation is entangled with the measured system and the environment.  Such a state does not appear to represent the agent seeing a definite and unique outcome. And yet, experience suggests that she always does.

This tension between unitary evolution and a common-sense understanding of measurement has been long known about, but only recently have  \textit{logical contradictions} been uncovered \cite{Frauchiger2018quantum, brukner2018no, Nurgalieva_2019, pusey2016qbism, healey2018quantum, allard2021no, bong2020strong, haddara2022possibilistic}. 
A particularly enlightening no-go theorem is found in \cite{bong2020strong}. There, the authors showed that quantum theory is incompatible with a theory-independent principle called `Local Friendliness'. This principle involves a commitment to observed measurement outcomes and settings being \textit{absolute} (that is, unique and non-relational), and to some weak locality assumptions. The Local Friendliness theorem therefore strongly suggests that quantum theory either forces a breakdown in the absoluteness of our observations, or else entails a strong form of nonlocality.



But there is a loophole. The theorem assumes the rules of standard probability theory can be used to translate its metaphysical assumptions into mathematical inequalities. 
If one rejects those rules, then one evades the no-go theorem and its consequences.
This is reminiscent of how, before an inequality-free proof of nonlocality was devised by Greenberger, Horne, and Zeilinger \cite{greenberger1990bell}, one could conceivably maintain `local realism' in spite of Bell's theorem by rejecting Bell's use of the probability calculus.

Now, this loophole in the Local Friendliness theorem was closed in \cite{haddara2022possibilistic}, which derived an inequality-free version. However, in order to avoid the use of the probability calculus, the new theorem made use of a `possibility calculus'. This calculus was provided by modal logic with Kripke structures. If one rejects the applicability of that framework, and the inferences between statements about possibilities that it licences, then one can again evade the no-go theorem.
This might even seem like the natural course of action, since it can easily be argued that the possible worlds encapsulated by Kripke structures are `too classical' for a quantum setting \cite{Nurgalieva_2019, haddara2022possibilistic}. 

Here, we show that this strategy fails. We provide a new no-go theorem demonstrating that the problem of absoluteness cannot be solved by rejecting or modifying modal logic alone.  We do so by demonstrating a contradiction between absoluteness and quantum theory that does not require any use of modal logic, or indeed the probability calculus. 

For this contradiction, one needs to assume a slightly stronger version of quantum theory than the one from \cite{bong2020strong, haddara2022possibilistic}, but one which is motivated by relativity. The idea is that one can take a Heisenberg cut along a simultaneity surface in any inertial reference frame to provide Born-rule predictions for pairs of space-like separated measurements. Thus the theorem reveals a three-way tension between (i) treating measurements unitarily (ii) treating all inertial reference frames as equally valid and (iii) an absolute observed reality. Given a version of quantum theory that models measurements unitarily and which fits naturally with special relativity, it simply cannot be true that there are absolute observed outcomes.


\section{The no-go theorem}

When a rational, competent agent believes they have witnessed the outcome of a measurement, by, for example, seeing a pointer move in a particular direction, we assume that there has been an `observed event'. One could, however, say that (i) there was no \textit{unique} event, perhaps because the universe split into different worlds where different events were observed; and/or that (ii) the event only happened relative to some particular perspective, such as the point of view of the observer, or the context of measurement. The assumption of the absoluteness of observed events rules out both (i) and (ii). In \cite{haddara2022possibilistic}, it was formulated as follows.
\begin{definition}[Absoluteness of observed events (AOE).]
Every observed event is an absolute single event, not relative to anything or anyone.
\end{definition}

We take this to have the following implication. If $n$ rational, competent agents each believe they have witnessed the outcome of a measurement, then there is a single, non-relative fact about the set of outcomes 
\begin{equation}
o=\{x_i \}_{i=1}^n  \in \{X_i \}_{i=1}^n    
\end{equation}
that were observed. Above, each $X_i$ denotes the set of possible outcomes of the $i$th measurement.

We now describe a version of quantum theory that is incompatible with this assumption. In this theory, the joint probabilities for a set of measurements performed at the same coordinate time $t$ are calculated by assuming the quantum state evolved unitarily up until time $t$, and then applying the Born rule. This is true even if prior to $t$ the system that is to be measured includes observers that take themselves to be performing measurements -- in this case, their measurements are treated like any other unitary interaction.

Crucially, we further assume that the above holds for any inertial reference frame. We summarize the target theory of our no-go theorem as follows.
\begin{definition}[Frame-independent quantum theory (FIQT).]
 A version of quantum theory in which predictions for measurement outcomes are made in the following way:
 \begin{enumerate}[(i)]
     \item the state of a quantum system is treated as evolving unitarily prior to the measurements whose outcomes we want to predict, even if that state includes observers;
     \item joint probabilities for measurements performed at the same coordinate time are then given by the Born rule; and
     \item all this holds in any inertial frame of reference.
 \end{enumerate}
\end{definition}

Finally, we can state our result.

\begin{theorem}[No-go result.] AOE and FIQT are not both correct. \end{theorem}

\section{Proof}

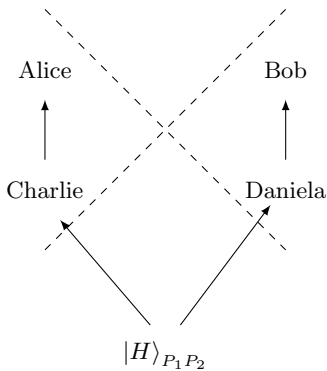
\begin{figure} 
    
   \begin{tikzpicture} [scale=0.2]
	\begin{pgfonlayer}{nodelayer}
		\node [style=none] (0) at (-0.25, -2.25) {};
		\node [style=none] (1) at (0, -2.75) {$\ket{H}_{P_1P_2}$};
		\node [style=none] (2) at (-2, 0) {};
		\node [style=none] (3) at (-2, 0) {Charlie};
		\node [style=none] (4) at (2, 0) {};
		\node [style=none] (5) at (2, 0) {Daniela};
		\node [style=none] (6) at (-2, 2) {};
		\node [style=none] (7) at (-2, 2) {Alice};
		\node [style=none] (8) at (2, 2) {Bob};
		\node [style=none] (9) at (-1.75, -0.5) {};
		\node [style=none] (10) at (0.25, -2.25) {};
		\node [style=none] (11) at (1.75, -0.25) {};
		\node [style=none] (12) at (2, 0.5) {};
		\node [style=none] (13) at (2, 1.5) {};
		\node [style=none] (14) at (-2, 0.5) {};
		\node [style=none] (15) at (-2, 1.5) {};
		\node [style=none] (16) at (-2, -1) {};
		\node [style=none] (17) at (2, 3) {};
		\node [style=none] (18) at (2, -1) {};
		\node [style=none] (19) at (-2, 3) {};
	\end{pgfonlayer}
	\begin{pgfonlayer}{edgelayer}
		\draw [style=arrow plain] (0.center) to (9.center);
		\draw [style=arrow plain] (10.center) to (11.center);
		\draw [style=arrow plain] (12.center) to (13.center);
		\draw [style=arrow plain] (14.center) to (15.center);
		\draw [style=dashed] (16.center) to (17.center);
		\draw [style=dashed] (19.center) to (18.center);
	\end{pgfonlayer}
\end{tikzpicture}
    \caption{The scenario that proves the no-go theorem. Charlie and Daniela each measure one qubit, before Alice and Bob each perform a supermeasurement on one agent and one qubit. The dotted lines represent light-like rays.}
    \label{fig:causal_structure}
\end{figure}

As in \cite{haddara2022possibilistic}, the present result will be proven with an adaptation of Hardy's paradox \cite{hardy1992quantum, hardynonlocality1993}.\footnote{The connection between Hardy's paradox and paradoxes involving supermeasurements has been noted in \cite{pusey2018inconsistent, brukner2018no, aaronson2018hard}.}
However, the scenario we will consider will not feature any choices of measurement (and it is in that sense closer to the original Frauchiger-Renner setup \cite{Frauchiger2018quantum}, and the one discussed in \cite{pusey2016qbism, healey2018quantum, baumann2019comment, healey2019reply}). It involves four measurements, two of which are supermeasurements, as illustrated in Figure \ref{fig:causal_structure}.

We begin with a pair of particles prepared in the `Hardy state'
\begin{equation}
    \ket{H} = \frac{1}{\sqrt{3}}(\ket{00}_{P_1P_2}+\ket{01}_{P_1P_2}+\ket{10}_{P_1P_2})
\end{equation}
The agents Charlie and Daniela each have access to one of the particles, $P_1$ and $P_2$, respectively. They each perform a $Z$ basis measurement, which can be described unitarily as a CNOT interaction,
where the target is a qubit representing the agent's memory.\footnote{Of course, a more realistic model of the measurement would not represent the memory as a qubit, and would account for the role of the environment. But as long as the measurement is still treated unitarily, essentially the same argument applies.} 

Upon initializing the memory in the `ready' state $\ket{0}$, the CNOT gives rise to an isometry $U$ that coherently copies the qubit in the $Z$ basis onto the memory. For example, Charlie's measurement isometry is defined as $U\ket{i}_{P_1}=\ket{i}_{C} \ket{i}_{P_1}$ for $i \in \{0, 1 \}$. 

After these initial measurements, Alice and Bob each have access to a composite quantum system consisting of a memory and a particle, $C \otimes P_1$ and $D \otimes P_2$ respectively.
They each measure a basis containing the states $U\ket{\pm}$, obtained by applying the measurement isometry from above to the $X$ basis states $ \ket{\pm}:=\frac{\ket{0} \pm \ket{1}}{\sqrt{2}}$ of the original particle. These supermeasurements can be thought of as first `undoing' Charlie's and Daniela's measurements, and then measuring the $X$ bases for the original particles.

Importantly, we assume that both Alice and Charlie are space-like separated from both Bob and Daniela. This is illustrated in Figure \ref{fig:causal_structure}, in which the dotted lines represent light-like surfaces.

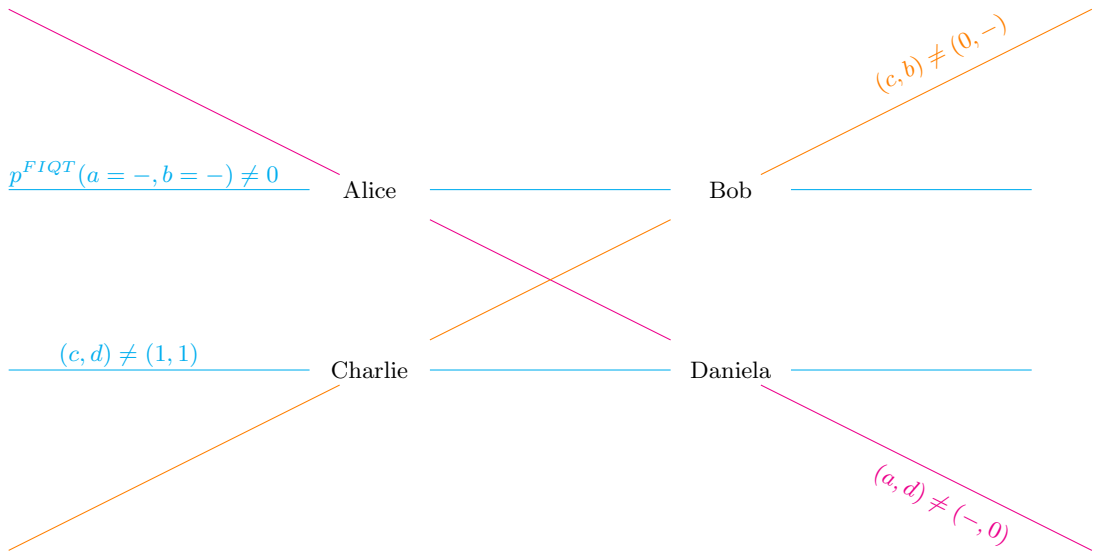
\begin{figure*}
    \centering
    \begin{tikzpicture}[scale=0.2]
	\begin{pgfonlayer}{nodelayer}
		\node [style=none] (0) at (-3, -1.5) {};
		\node [style=none] (1) at (-3, -1.5) {Charlie};
		\node [style=none] (2) at (3, -1.5) {};
		\node [style=none] (3) at (3, -1.5) {Daniela};
		\node [style=none] (4) at (-3, 1.5) {};
		\node [style=none] (5) at (-3, 1.5) {Alice};
		\node [style=none] (6) at (3, 1.5) {Bob};
		\node [style=none] (7) at (-2, -1.5) {};
		\node [style=none] (8) at (2, -1.5) {};
		\node [style=none] (9) at (-2, 1.5) {};
		\node [style=none] (10) at (2, 1.5) {};
		\node [style=none] (11) at (-2, 1) {};
		\node [style=none] (12) at (2, -1) {};
		\node [style=none] (13) at (-3.5, 1.75) {};
		\node [style=none] (14) at (3.75, -1.75) {};
		\node [style=none] (15) at (-9, 4.5) {};
		\node [style=none] (16) at (9, -4.5) {};
		\node [style=none] (17) at (3.5, -1.75) {};
		\node [style=none] (18) at (-4, 1.5) {};
		\node [style=none] (19) at (-9, 1.5) {};
		\node [style=none] (20) at (4, 1.5) {};
		\node [style=none] (21) at (8, 1.5) {};
		\node [style=none] (22) at (-4, -1.5) {};
		\node [style=none] (23) at (-9, -1.5) {};
		\node [style=none] (24) at (4, -1.5) {};
		\node [style=none] (25) at (8, -1.5) {};
		\node [style=none] (26) at (2, 1) {};
		\node [style=none] (27) at (-2, -1) {};
		\node [style=none] (28) at (3.5, 1.75) {};
		\node [style=none] (29) at (9, 4.5) {};
		\node [style=none] (30) at (-9, -4.5) {};
		\node [style=none] (31) at (-3.5, -1.75) {};
		\node [style=none] (32) at (-6.75, 1.75) {\color{cyan} $p^{FIQT}(a=-, b=-) \neq 0$};
		\node [style=none] (33) at (-7, -1.25) {$\color{cyan} (c, d) \neq (1, 1)$};
		\node [style=none, rotate=-26.77] (34) at (6.5, -3.75) {$\color{magenta} (a, d) \neq (-, 0)$};
		\node [style=none, rotate=26.77] (35) at (6.5, 3.75) {\color{orange}$(c, b) \neq (0, -)$};
		\node [style=none] (36) at (2, -1) {};
		\node [style=none] (37) at (-2, 1) {};

	\end{pgfonlayer}
	\begin{pgfonlayer}{edgelayer}
		\draw [style=cyan] (7.center) to (8.center);
		\draw [style=cyan] (9.center) to (10.center);
		\draw [style=magenta] (11.center) to (12.center);
		\draw [style=magenta] (15.center) to (13.center);
		\draw [style=magenta] (17.center) to (16.center);
		\draw [style=cyan] (18.center) to (19.center);
		\draw [style=cyan] (22.center) to (23.center);
		\draw [style=cyan] (20.center) to (21.center);
		\draw [style=cyan] (24.center) to (25.center);
		\draw [style=orange] (26.center) to (27.center);
		\draw [style=orange] (29.center) to (28.center);
		\draw [style=orange] (31.center) to (30.center);
	\end{pgfonlayer}
\end{tikzpicture}
    \caption{The argument summarized diagrammatically. Each straight line represents a surface of simultaneity in some frame corresponding to the color. Three simultaneity surfaces are annotated with the pair of purportedly absolute outcomes they rule out, assuming the predictions of FIQT are accurate. These three facts together imply $(a, b) \neq (-, -)$, and yet the FIQT prediction for this pair of outcomes using the blue frame $p^{FIQT}(a=-, b=-) \neq 0$.}
    \label{fig:ref_frames}
\end{figure*}

This scenario features four rational, competent agents, who each believe they have witnessed a unique measurement outcome. AOE states that each agent really did witness a unique outcome, and the fact about which outcome they observed is not relative to anything or anyone. These unique and non-relational outcomes may be written as a tuple:
\begin{equation}
    o = (a, b, c, d)
\end{equation}
where $a$ represents Alice's outcome, $b$ represents Bob's outcome, and so on.

If FIQT and AOE are both correct, then presumably we can use FIQT to make predictions about $o$. In particular, if FIQT allows us to calculate a joint probability of zero for some possible values $(i, j)$ of two out of the four outcomes (e.g. $p^{FIQT}(c=i, d=j)=0$), then we assume that those are not the actual outcomes in $o$ (in this case, $(c, d) \neq (i, j)$). On the other hand, if FIQT predicts a nonzero probability for some pair of outcomes, then we assume that, given enough runs of the experiment, there will be at least one run in which this pair is observed.

We will now show that when we assume AOE, FIQT makes self-contradictory predictions. The argument, illustrated in Figure \ref{fig:ref_frames}, proceeds by combining the FIQT predictions made about  $o$ in various reference frames to rule out a particular pair of values for $a$ and $b$, which FIQT nevertheless assigns a non-zero probability.\footnote{Thus, the reference frames here play a similar role to the rules of modal logic and the `Possibilisitic Local Agency' assumption in \cite{haddara2022possibilistic}. Namely, they justify the chaining together of statements associated with different measurement contexts in order to derive a contradiction.}

We start by considering the blue frame from Figure \ref{fig:ref_frames}, in which Charlie and Daniela measure simultaneously. FIQT allows us to compute a joint probability of 0 for both agents to observe the outcome corresponding to the $\ket{1}$ state, which we denote `1'
\begin{equation}
    p^{FIQT}(c=1, d=1) = |\bra{11}\ket{H}|^2 = 0
\end{equation}
We infer that Charlie and Daniela do not both observe `1'
\begin{equation} \label{obs1}
    (c, d) \neq (1, 1)
\end{equation}

We can also consider the pink frame, where Alice and Daniela measure at the same time. We use FIQT to compute the joint probability for Alice to see the outcome `-' corresponding to $U\ket{-}$ and Daniela to see `0' corresponding to $\ket{0}$. We do this by treating Charlie's measurement as a unitary interaction (which gives the isometry $U$ when we prepare Charlie's memory in the appropriate state). We get
\begin{equation}
\begin{split}
    &p^{FIQT}(a=-, d=0) \\ & \ = |\bra{-}_{P_1}\bra{0}_{P_2}(U^\dag \otimes I)(U \otimes I)\ket{H}_{P_1P_2}|^2 \\
    & \ = |\bra{-}_{P_1}\bra{0}_{P_2}\ket{H}_{P_1P_2}|^2 \\
    & \ = 0
\end{split}
\end{equation}
meaning that this pair of observations did not take place
\begin{equation} \label{obs2}
    (a, d) \neq (-, 0)
\end{equation}

In the orange frame,  Charlie and Bob measure simultaneously. Repeating the argument above, we conclude that 
\begin{equation} \label{obs3}
    (c, b) \neq (0, -)
\end{equation}
From (\ref{obs1}), (\ref{obs2}), and (\ref{obs3}), it logically follows that 
\begin{equation} \label{obs3}
    (a, b) \neq (-, -)
\end{equation}
meaning that Alice and Bob never both see a `-' outcome, no matter how many times we run the experiment. 

Finally, we return to the blue frame, where Alice and Bob measure simultaneously, and compute the FIQT prediction by treating both Charlie's and Daniela's measurements unitarily. We get
\begin{equation}
    \begin{split}
        &p^{FIQT}(a=-, b=-)  \\ & \ = |\bra{-}_{P_1}\bra{-}_{P_2}(U^\dag \otimes U^\dag)(U \otimes U)\ket{H}_{P_1P_2}|^2 \\
        & \ =|\bra{-}_{P_1}\bra{-}_{P_2}\ket{H}_{P_1P_2}|^2 \\
        & \ = 1/12 \neq 0
    \end{split}
\end{equation}

Therefore, FIQT predicts that Alice and Bob \textit{do} sometimes both see the `1' outcome, contradicting our earlier conclusion. Assuming both AOE and FIQT has forced us to make two contradictory predictions about what Alice and Bob observe. It follows that measurements cannot be treated both unitarily and in a frame-independent way in any world where observed measurement outcomes are absolute. \qed

\section{Related results}

Some readers might prefer a proof of the no-go theorem in which FIQT predicts that the forbidden result \textit{always obtains}, rather than just $1/12$ of the time. Such a result is possible via a similar adaptation of Mermin's version \cite{mermin1990quantum} of the GHZ nonlocality argument \cite{greenberger1990bell}. To prove the result, one assumes that Charlie, Daniela, and Ethan are arranged in a triangle in space, and perform the  $X$ measurements from \cite{mermin1990quantum}. Then  Alice, Bob, and Fiona are arranged in the same triangle but at a later time, and perform supermeasurements that are effectively the $Y$ measurements from \cite{mermin1990quantum}.

Others might prefer the frame-independence assumption to be stated theory-independently, rather than absorbed into our definition of quantum theory.
To do this, one can stipulate that if a pair of outcomes never occurs in some experiment $E$, then the same pair of outcomes also never occurs in another experiment $E'$, which is identical to $E$ in every respect except that the experimental setup in $E'$ is in motion relative to the setup in $E$. Then, once one additionally assumes that quantum theory holds in at least one reference frame, one gets a composite assumption similar to FIQT. One can prove that this composite assumption is inconsistent with AOE via a similar argument to the one above, except that rather than three reference frames, one considers three duplicates of the same experiment, each in motion relative to the each other.

\section{Discussion}

Two of the deepest principles of modern physics are those of unitary evolution and the equivalence of reference frames. Yet here we have seen clearly that their combination rules out a principle that many would consider even more fundamental: the absoluteness of observed events. And the proof cannot be escaped by revisions to the probability calculus, or even modal logic.

There are no conservative responses to this. Either one rejects FIQT, and with it, a central principle of at least one of the two pillars of modern physics. Or, one rejects AOE, and embraces a reality in which the objectivity of data seems to be ruled out in principle; a world in which even our observations are somehow fragmented into worlds or perspectives or contexts, which cannot be glued back together to tell a unique and consistent story.

This raises the formidable question of which radical response to adopt. Various options have been discussed elsewhere (for example, in \cite{Frauchiger2018quantum, Nurgalieva_2019, Cavalcanti_2021}), and we won't have anything more to say about them here. But it also raises another question. Namely, precisely which features of quantum theory lead to the paradox described above? Or, on a closely related note, which theories in general lead to a breakdown in the absoluteness of observed events, through a generalization of the argument here?\footnote{We already know that it is not \textit{just} quantum theory: \cite{vilasini2019multi} describes a similar paradox arising from PR boxes \cite{popescu1992generic}.}  

Beyond helping us understand more deeply the origin of the problem with absoluteness, answering these questions would help us judge whether future theories of physics -- for example a fully-fledged theory of quantum gravity -- will also be inconsistent with absoluteness. And they are particularly pertinent questions to ask in the context of the current no-go theorem, which is proven with a very simple argument, and which is therefore amenable to generalization. Upcoming work will provide such a generalization, by developing the generalization of the Heisenberg cut introduced in \cite{vilasini2019multi}.

\section*{Acknowledgements}

We are pleased to thank Nicola Pinzani, Vilasini Venkatesh, Marwan Haddara, Tein Van Der Lugt, Maria Violaris, Richard Howl, Augustin Vanrietvelde, and Hl\'er Kristj\'ansson for some very helpful discussions.
We also thank Alexandra Elbakyan for her help to access the scientific literature. NO acknowledges funding from the UK Engineering and Physical Sciences Research Council (EPSRC). This publication was made possible through the support of the grant 61466 ‘The Quantum Information Structure of Spacetime (QISS)’ (qiss.fr) from the John Templeton Foundation. The opinions expressed in this publication are those of the authors and do not necessarily reflect the views of the John Templeton Foundation.

\bibliography{refs}

\end{document}